\documentclass[aps,prd,showpacs,preprint,tightenlines]{revtex4}
\usepackage{graphicx,amsmath}

\def \beq{\begin{equation}}         \def \eeq{\end{equation}}
\def \be{\begin{eqnarray}}
\def \ee{\end{eqnarray}}
\def \beqa{\begin{eqnarray}}
\def \eeqa{\end{eqnarray}}

\def \bea{\begin{array}}        \def \eea{\end{array}}

\def\nn{\nonumber \\ }

\def\a{\alpha} 
  
\def\g{\gamma} \def\G{\Gamma}
 \def\D{\Delta}
\def\m{\mu}    
\def\n{\nu}    
\def\p{\pi}    
 
\def \x{\xi}
\def \e{\epsilon}

\def \t{\tau}
\def \l{\lambda}

\def \abs#1{\left| #1 \right|}

\def \tr#1{\left[ #1\right]}

\def \gfive{\gamma_{5}}
\def \sh#1{#1\!\hskip-5pt /}

\def\mint#1{\int {d^d #1\over (2\pi)^d}\ }

\def\fint#1{\int^1_0 d#1\ }
\def\fintb#1#2{\int^1_0 d#1\int^{1-#1}_0 d#2 \ }

\newcommand{\ignor}[1]{}

\newcommand{\gmt}{$g-2$ }
\begin{document}

\title{Lepton flavor-changing Scalar Interactions and Muon $g-2$}
\author{Yu-Feng Zhou} \email[Email: ]{zhou@theorie.physik.uni-muenchen.de}
\affiliation{Ludwig-Maximilians-University Munich, \\
Sektion Physik. Theresienstra$\beta$e 37, D-80333. Munich, Germany}
\author{Yue-Liang Wu} \email[Email: ]{ylwu@itp.ac.cn}
\affiliation{ Institute of Theoretical Physics, Chinese Academy of
Science, Beijing 100080, China }
\date{\today}

 \begin{abstract} 
   A systematic investigation on muon anomalous magnetic moment and
   related lepton flavor-violating process such as $\m\to e\g$, $\t\to
   e\g$ and $\t\to \m\g$ is made at two loop level in the models with
   flavor-changing scalar interactions.  The two loop diagrams with
   double scalar exchanges are studied and their contributions are
   found to be compatible with the ones from Barr-Zee diagram.  By
   comparing with the latest data, the allowed ranges for the relevant
   Yukawa couplings $Y_{ij}$ in lepton sector are obtained.  The
   results show a hierarchical structure of $Y_{\m e, \t e} \ll Y_{\m
     \t} \simeq Y_{\m\m}$ in the physical basis if $\Delta a_{\mu}$ is
   found to be $>50\times 10^{-11}$.  It deviates from the widely used
   ansatz in which the off diagonal elements are proportional to the
   square root of the products of related fermion masses.  An
   alternative Yukawa coupling matrix in the lepton sector is
   suggested to understand the current data. With such a reasonable
   Yukawa coupling ansatz, the decay rate of $\t\to \m\g$ is found to
   be near the current experiment upper bound.  
\end{abstract}
\maketitle
%
%
\section{Introduction}
Recently, the Muon \gmt Collaboration at BNL reported their improved
result on the measurement of muon anomalous magnetic moment ( \gmt)
\cite{Bennett:2002jb}.  Combining with the early measurements in CERN and
BNL, the new average value of muon \gmt  is as follows
\be
a_{\mu}^{exp}=(116592030\pm 80 )\times 10^{-11}
\ee
This result confirmed the earlier measurement\cite{Brown:2001mg} with a much higher
precision. With this new result the difference between experiment and
the Standard Model~ (SM) prediction is enlarged again.  The most
recent analysis by different groups are given by 
\be
\D a_{\m} \equiv a^{exp}_{\m}-a^{SM}_{\m}
=\left\{ 
\begin{array}{ll}
           (303.3 \pm106.9)\times 10^{-11} & \mbox{\cite{FJ02}} \\
           (297.0 \pm 107.2 )\times 10^{-11} (ex)& \mbox{ \cite{HMNT}} \\
           (357.2 \pm 106.4)\times 10^{-11}(in) &\mbox{ \cite{HMNT}}
\end{array}
\right.
\ee

As the large muon \gmt  may imply the existence of new physics beyond the
SM, in the recent years large amount of work has been done in checking the
new physics contributions to muon \gmt by using model
dependent\cite{models-g2,wu:2001vq} and independent approaches
\cite{Raidal:2001pf}. 

In this work, we would like to focus on a general discussion on the
models with lepton flavor-changing scalar interactions where the new
physics contributions mainly arise from additional Yukawa couplings.  Such models
may be considered as the simple extension of the standard model (SM)
with more than one Higgs doublet $\phi_i \ (i > 1)$ but without
imposing any discrete symmetry. For example  the extension of
SM with two Higgs doublets (S2HDM)\cite{wu:S2HDM} motivated from the
spontaneous CP violation\cite{Lee:1973iz,Lee:1974jb}. 

%

The general form of Yukawa interaction  reads
\be
{\mathcal L}_{Y}=\bar{\psi}^{i}_{L} Y^{a}_{ij}  \psi^{j}_{R}\phi_{a} ,
\label{Yukawa-interaction}
\ee
where $Y^{a}_{ij}$ ($i,\ j =1,2,3$) is the Yukawa coupling matrices.
The index $a=1,2, \cdots$ labels the Higgs doublets. The behavior of
the Yukawa interactions depends on the texture of Yukawa coupling
matrices. In general there are two kind of ansatz on Yukawa coupling
matrices in mass eigenstates:
1. Yukawa coupling matrices of the scalar interactions are diagonal due to 
some discrete global symmetry \cite{2HDM-type12}.
2. Yukawa coupling matrices  contain
non-zero off-diagonal elements which are naturally suppressed by the
light quark masses \cite{cheng:1987rs,sher:1991km}. 
In the following sections (section {\bf I$\!$I} and {\bf I$\!$I$\!$I})
we discuss at two loop level the constraints on those Yukawa coupling
matrix elements under the above two ansatz and will mainly focus on
the latter one.  One kind of two loop diagrams with double scalar
exchanges are studied in detail and their contributions to muon \gmt
are found to be compatible with the one from Barr-Zee diagram.  In
section {\bf I$\!$V}, combined constraints from muon \gmt and several
lepton flavor violating (LFV) processes are obtained. We note that unlike other
experiments which often impose upper bounds of parameters in the new
physics models, the current data on muon $g-2$ may provide nontrivial
lower bounds. It is founded that a small lower bound of  $\Delta
a_{\mu} >  50\times 10^{-11}$ will significantly modify the texture
of Yukawa coupling matrix and make it deviate from the widely used
ansatz in which the off diagonal elements are proportional to the
square root of the products of related fermion masses.  




\section{Muon \gmt from diagonal Yukawa couplings}
\begin{figure}[htb]\begin{center}
\includegraphics[width=0.6\textwidth]{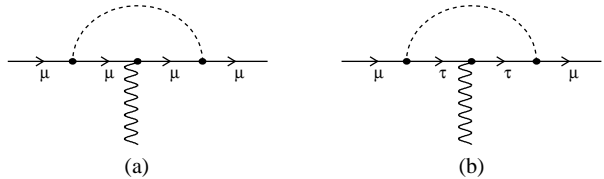}
\caption{One loop  diagram contribution to  muon \gmt. The dashed curves represent
the scalar or pseudo-scalar propagator. 
(a)  Flavor conserving Yukawa interactions.
(b) Flavor changing Yukawa interactions in which $\m$ changes into $\t$  in the
loop.  }
\label{1l.eps}
\end{center}\end{figure}

The ansatz of zero off-diagonal matrix elements is often used to avoid
the flavor-changing neutral current (FCNC) at tree level which was
originally suggested from the kaon physics, such as $K\rightarrow
\mu^+ \mu^-$ decay and $K^{0}-\overline{K}^{0}$ mixing.  Such a
texture structure of the Yukawa couplings can be obtained by imposing
some kind of discrete symmetries\cite{2HDM-type12}.  The minimal SUSY
Standard Model (MSSM) and the two-Higgs-doublet model (2HDM) of type
{\bf I} and {\bf I$\!$I} can be cataloged into this type. In such
models, the Yukawa interactions are flavor conserving and the
couplings are proportional to the related fermion masses
\be
Y_{ii}&=&{\Large { g m_{i}\over 2 m_{W} }} \xi_{i}  \quad {\rm and} \quad  Y_{ij}=0.
 \ (i \neq j)
\ee
where $g$ is the weak coupling constant and $m_W$ is the mass of $W$ boson.
$\xi_{i}$ is the rescaled coupling constant. In the minimal SUSY model
and the 2HDM of type {\bf I$\!$I},  $\xi_i=\tan\beta (\cot\beta)$  for down (up) type fermions.
%
The corresponding Feynman diagrams contributing to muon \gmt at one
loop level which is shown in Fig.\ref{1l.eps}a , which have been recently
discussed and compared with the current data in
Ref.\cite{Dedes:2001hh,Dedes:2001nx,Krawczyk:2001pe}.

As the muon lepton mass is small, i.e., $m_{\m}\ll m_{\phi}$,
where $m_{\phi}$ is the mass of scalar ($\phi=h$)  or pseudo-scalar ($\phi = A$),  the one loop
contribution to muon \gmt  can be written as \cite{oneloop-g2}
\be
\D a_{\m}=\pm{ 1\over 8\p^{2}} { m^2_{\m}\over m^2_{\phi}}
\ln \left( {m^2_{\phi}\over m^2_{\m}}\right) Y^2_{ii}
\ee
where the sign ``+ (-)'' is for scalar ($\phi=h$) (pseudo-scalar $\phi=A$) exchanges.
It can be seen from the above equation that the one loop scalar
contribution is not large enough to explain the current data.
Even for a large value of $\xi_\mu=\tan\beta\sim 50$, one still needs a very
light mass of the scalar $M_h\sim 5 $ GeV,
which seems not favored by the LEP experiment. The situation will be even
worse when both the scalar and pseudo-scalar are included  as their
contributions have opposite signs.
%

\begin{figure}[htb]\begin{center}
\includegraphics[width=0.6\textwidth]{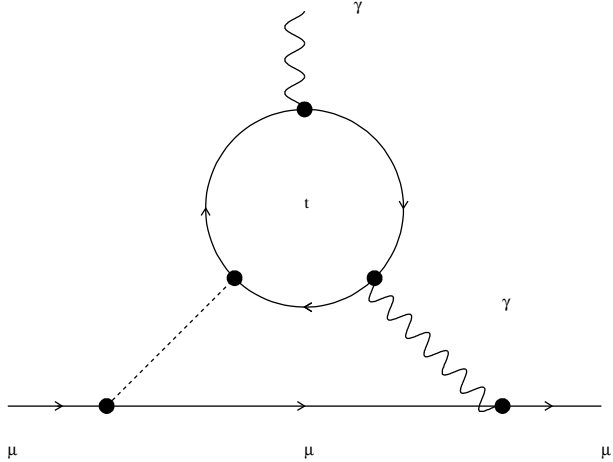}
\caption{Two loop Barr-Zee  diagram contribution to muon \gmt. 
  }
\label{barrzee.eps}
\end{center}
\end{figure}

The situation may  be quite different if one goes to  two loop level.
From the well known Barr-Zee mechanism\cite{BarrZee} ( see. Fig.\ref{barrzee.eps} )in which the
scalar or pseudo-scalar couples to a heavy fermion loop.
As the Yukawa couplings are no longer suppressed by the light
fermion mass,  the two loop contributions
could be considerable. Taking the top quark loop as an example, the
two loop Barr-Zee diagram contribution  to muon \gmt is given by
\be\label{BarrZeeG2}
\Delta a^h_\mu &=& { N_c q^2_t  \over \pi^2 }
{m_\mu m_{t} \over m^2_{\phi}}  F\left({ m^2_{t}\over m^2_{\phi} }\right)
Y_{tt} Y_{\m\m}
\label{BZ}
\ee
where $N_c=3$ and $q_t=2/3$ are the color number and  the charge of top quark respectively.
The integral function $F(z)$ has the following form\cite{BarrZee}
\be
F(z)=
\left\{
\begin{array}{cc}
-{1\over2}\int^1_0 dx {1-2x(1-x)\over x(1-x)-z} \ln{x(1-x)\over z}
& \text{for scalar} \\
{1\over2}\int^1_0 dx {1\over x(1-x)-z}\ln {x(1-x)\over z}
& \text{for pseudo-scalar}
\end{array}
\right.
\nn
\ee

\begin{figure}[htb]\begin{center}
\includegraphics[width=0.6\textwidth]{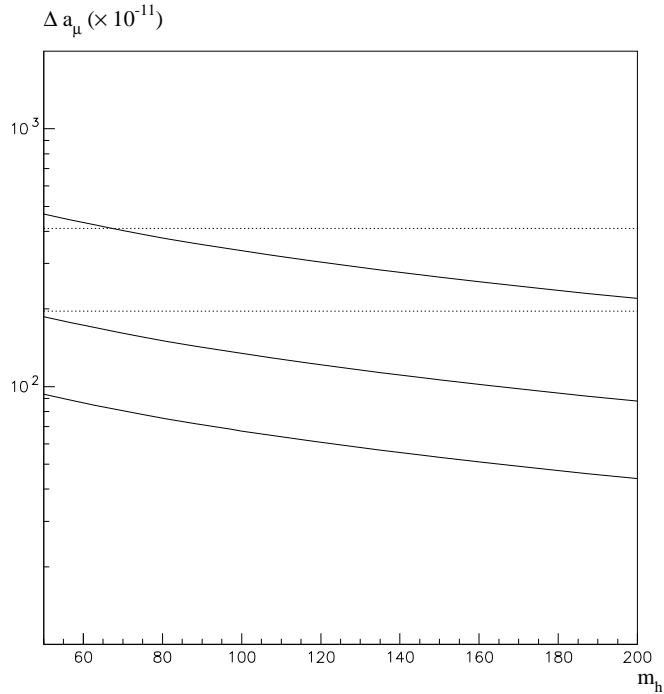}
\caption{Contribution to muon \gmt from the two
loop   Barr-Zee diagrams. The three solid curves (from down to up) correspond to
 $Y_{\m \m}(\xi_\m)=2\times 10^{-2}(48.3)$, $4\times 10^{-2}(97.6)$ and $1\times 10^{-1}(273.9)$ respectively.
The  horizontal lines represent the $1\sigma$ allowed rang 
from Ref.\cite{FJ02}
}

\label{barrzee-g2.eps}
\end{center}\end{figure}

It is noticed that the contributions from Barr-Zee diagram through
scalar and pseudo-scalar exchanges have also different sign, negative
for scalar and positive for pseudo-scalar, which is just opposite to
the one loop case.  Thus there exists a cancellation between one and
two loop diagram contributions. It was found in
Refs.\cite{Chang:2000ii,Cheung:2001hz} that the pseudo-scalar
exchanging Barr-Zee diagram can overwhelm its negative one loop
contributions and results in a positive contribution to \gmt.  For a
sufficient large value of the coupling $\xi_\mu=\tan\beta\sim 50$, its
contribution can reach the $2\sigma$ experimental bound with $m_{\phi} \leq
70$GeV.  To avoid the cancellation between scalar and pseudo-scalar
exchange, the mass of the scalar boson has to be pushed to be very
heavy ( typically greater than 500 GeV).  In Fig.\ref{barrzee-g2.eps} 
the numerical calculation of Barr-Zee diagram contribution to muon \gmt is presented,
which agrees with those results.

%


%
%
%

\section{Muon \gmt from off-diagonal Yukawa couplings}

When imposing the strict discrete symmetries to Yukawa interaction,
the off-diagonal elements of Yukawa coupling matrix are all zeros. This
is the simplest way to prevent the theory from tree level
FCNC. However, to meet the constraints from the data on
$K^0-\overline{K}^0$ mixing and $K\to \m^{+}\m^{-}$ the off-diagonal
elements do not necessarily to be zero. An alternative way is to
impose some approximate symmetries such as global family symmetry
\cite{wu:S2HDM} on the Lagrangian. This results in the second ansatz of
the Yukawa matrices in which  small off-diagonal matrix elements are allowed,
which  leads to an enhancement for many flavor changing processes.
As the constraints from $K^{0}-\overline{K}^{0}$ mixing are strong, the corresponding
off-diagonal matrix elements should be very small. However, up to now there is no
such strong experimental constraints on the FCNC processes involving heavier flavors such as $c$ and $b$ quarks.
The possibility of off-diagonal elements associated with the second and the third
generation fermions are not excluded.
One of the widely used ansatz of the Yukawa
matrix basing  on the hierarchical fermion mass spectrum
$m_{u,d}\ll m_{c,s}\ll m_{t,b}$ was proposed by Cheng and Sher \cite{cheng:1987rs,sher:1991km}.
In this ansatz, the off-diagonal matrix element has the following form:
\be
Y_{ij}={g\sqrt{m_{i}m_{j}} \over 2 m_{W}} \xi_{ij}
\label{Cheng-Sher_Ansartz}
\ee
where  $\xi_{ij}$s are the rescaled Yukawa couplings which are roughly of the same order of
magnitudes for all  $i,j$s.  In this ansatz, the scalar or pseudo-scalar mediating $d-s$
transition is strongly suppressed by small  factor
$\sqrt{m_{d}m_{s}}/(2 m_W)\simeq 4\times 10^{-4}$, which easily
satisfies the constraints from $\D m_{K}, \e_{K}$ and $\G(K\to \m^{+}\m^{-})$.  As the
couplings grow larger for heavier fermions, the tree level FCNC processes may give
considerable contributions in $B^0-\bar{B}^0$ mixing, $\m^{+}\m^{-}\to t c,\m\t$
and several rare $B$ and $\t$ decay modes\cite{Atwood:1995ej,Atwood:1997vj,wu:1999fe,sher:2000uq}.

Unlike the flavor-conserving one loop diagrams, the flavor-changing
one loop diagrams (see Fig.\ref{1l.eps}b) with internal heavy fermions can give large
contribution to muon \gmt.  The reason is that the  loop integration
yield an enhancement factor of
$\sim m_{i}\ln(m^2_{i}/m^2_{\phi})/(m_{\m} \ln(m^2_{\mu}/m^2_{\phi}))$.
  For the internal $\t$ loop, it is a factor of
${\mathcal{O}}(10)$.  If one uses the scaled coupling $\xi_{\m\t}$ and
takes $\xi_{\m\t}\simeq \xi_{\m}$ as in the ``Cheng-Sher'' ansatz,
the value of the enhancement factor can reach ${\mathcal{O}}(10^2)$.
In the following discussion, for simplicity we only take one loop diagram with
internal $\tau$ loop into consideration as it is dominated over other fermion loops
in the case  that the Yukawa couplings are of  the same order of magnitudes. 

The expression of one loop flavor changing diagram contribution to muon
\gmt is given by \cite{nie:1998dg}
\be
\Delta a_\mu
&=&\pm{1 \over 8 \pi^2} {m_\mu m_\tau  \over  m^2_{\phi}}
     \left(\ln {m^2_{\phi}\over m^2_\tau}-{3\over2}\right)
     Y_{\mu\tau}^2
\ee
where the sign ``+ (-)'' is for scalar ($\phi=h$) (pseudo-scalar $\phi=A$) exchanges.
For detailed discussion on one loop flavor changing diagram, we refer to
the Refs.\cite{Kang:2001sq,Diaz:2000cm}.

As the two loop contribution to  muon \gmt is more considerable via the Barr-Zee
mechanism in flavor-conserving case, it is nature to go further to consider the
same diagram with flavor-changing couplings. However, in the case of muon \gmt,
as the initial and final states are all muons, it is easy to see that the Barr-Zee
diagram with flavor-changing coupling can not contribute.
%
%
It can only appear in flavor changing process such as $\m\to e\g$.  
The non-trivial two loop diagrams which give non-negligible contribution to \gmt are those
diagrams (as shown in Fig.\ref{2Ldh.eps}) which have two internal scalars with both of
them  coupling  to a heavy fermion loop.

\begin{figure}[htb]\begin{center}
\includegraphics[width=0.6\textwidth]{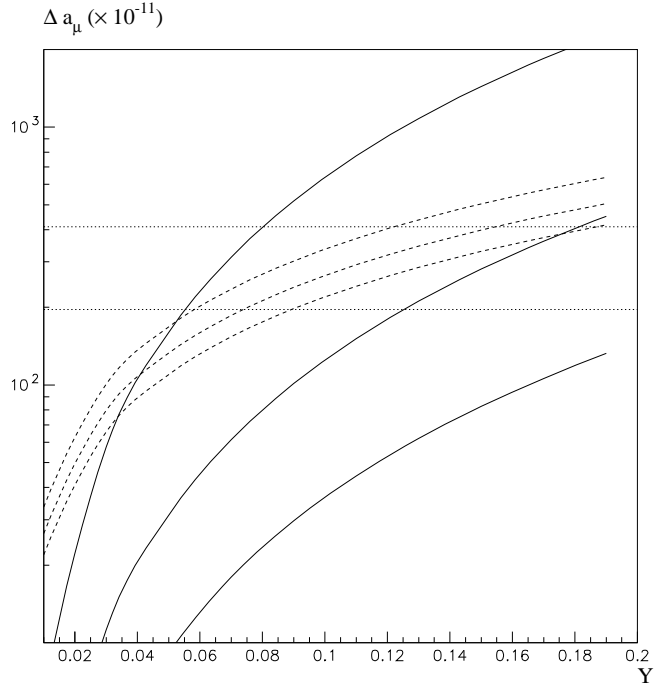} 
\caption{ Comparison between two loop Barr-Zee pseudo-scalar and 
double pseudo-scalar exchanging diagram in contribution to muon \gmt. The
contribution to muon \gmt is plotted as function of $Y=Y_{\m\t}=Y_{\m\m}$.
Three solid curves (from up to down) correspond to double scalar exchanging 
diagram contribution with scalar (pseudo-scalar) mass $m_{A}=100, 150, 200 $GeV  respectively.
Three dashed curves indicate the ones from two loop Barr-Zee diagrams with
pseudo-scalar exchange.   
The  horizontal lines represent the $1\sigma$ allowed range 
from Ref.\cite{FJ02}
 }
\label{twobarrzee-g2.eps}
\end{center}\end{figure}

It is known that large Yukawa couplings between scalar and heavy
fermions can compensate the loop suppressing factor $g^{2}/16\p^{2}$
and make the Barr-Zee diagram to be sizable. The same mechanism also
enhances the two-loop double scalar exchanging diagrams. Further more, in the
flavor changing case, the $\m$ lepton can go  into heavier lepton $\t$
in the lower loop, this may provide an additional enhancement in loop integration.
Taking the internal $t$-quark loop as an example, the ratio between the
contribution to muon \gmt from two-loop double scalar diagrams relative to the one from
Barr-Zee type diagrams can be roughly estimated by the
ratio between the couplings, which gives 
$\sim 
\xi_{t}\xi_{\m\t}^{2}  m_{t}m_{\t}   / ( 4\xi_{\m} m^2_{W} \sin^2\theta_{W}),
$ 
where $\theta_W$ is Weinberg angle with the value $\sin^2\theta_W\simeq 0.23$.
For the typical values of $\xi_{t}=1$ and $\xi_{\m\t}\simeq
\xi_{\m}=30$ the ratio is of order 1. Thus this kind of  two-loop double
scalar-exchanging diagram is compatible with the one of Barr-Zee type.
In the large  $m_{t}$ limit, the contribution to muon \gmt from two loop double
scalar (pseudo-scalar)  exchanging diagram has the following form
\be
\D a_{\m}
&=&
    \mp { N_{C}   m_{\t}m_{\m} m_{t}^2\over 16 \pi^{4}m_{\phi}^4} 
     \left( -{ 5\over 2}+\ln{ m_{\phi}^2\over m_{\t}^2 }\right)
     Y_{tt}^2 Y_{\m\t}^2
\label{tldh}
\nn
\ee
The details of the two loop calculations can be found in the appendix
at the end of this paper.  Comparing with the one-loop
flavor-changing diagram in the same way, one can see that the contribution from this
diagram could be  sizable .
%
%
%
%
%
For a comparison, the contribution to muon \gmt from two-loop double
pseudo scalar-exchanging diagrams and Barr-Zee diagrams with pseudo-scalar
are shown in Fig.\ref{twobarrzee-g2.eps}.

\begin{figure}[htb]\begin{center}
\includegraphics[width=0.6\textwidth]{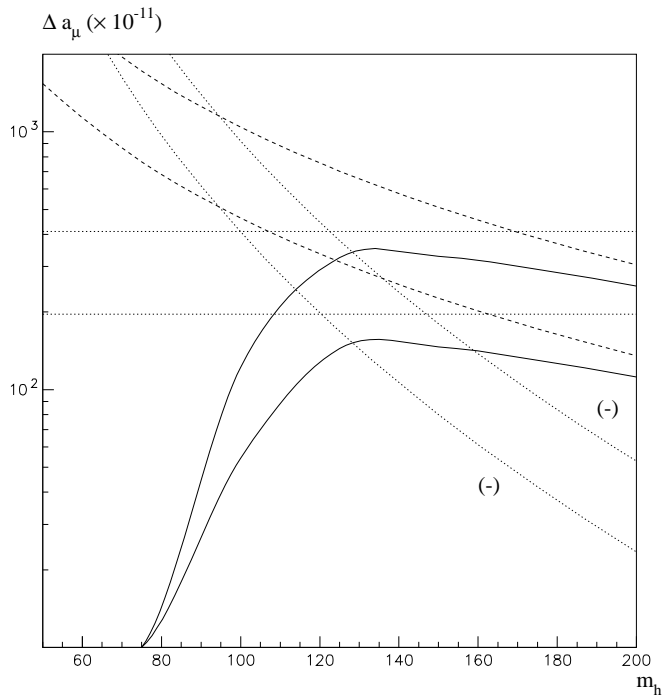} 
\caption{Comparison between one loop and two loop double scalar exchange
diagrams in contribution to muon \gmt. The contribution to muon \gmt is 
plotted as function of scalar mass. The two dashed curves represent  the contribution 
at  one loop with $Y_{\m\t}(\xi_{\m\t})$=0.12(70.6) (up) and 0.08(47) (down) respectively. The two
dotted curves correspond to the one from two loop double scalar diagram with the 
same couplings. (Note that their contribution are negative) The solid curves
are the total contribution to \gmt  from the both diagrams.
The  horizontal lines represent the $1\sigma$ allowed rang 
from Ref.\cite{FJ02}
   }
\label{onetwo-g2.eps}
\end{center}\end{figure}

To make the two kind of contributions comparable, we take 
$Y_{\m\m}=Y_{\m\t}\equiv Y$.
It can be seen  that the contribution from the former  highly depends on the coupling $Y$
and the scalar mass.  In the range $0.05\leq Y \leq 0.15$, the contribution from double
scalar-exchanging diagram is much large than the one from Barr-Zee diagram when
$m_{A}$ is about $100\sim 150$GeV. it decrease with $m_{A}$
increasing and becomes quite small when $m_{A}\sim 200$ GeV.
In Fig.\ref{onetwo-g2.eps}, the contribution to muon \gmt from two-loop double
scalar-exchanging diagrams is compared with the one from the corresponding
flavor-changing one loop diagrams. 
Note that just like the case of Barr-Zee diagram,
the two-loop double scalar-(pseudo-scalar) exchanging diagrams give
negative (positive)  contributions, which have the opposite signs as the one
from one loop scalar (pseudo-scalar)- exchanging diagram. The reason is 
that a closed fermion loop always contributes a minus sign.
It results in a strong cancellation between one and two loop diagram
contributions in the case of flavor changing couplings with real
Yukawa coupling constants.  The allowed range of the scalar mass will
be strongly constrained.  Taking $Y_{\m\t}=0.08(\xi_{\m\t}\simeq 50)$
, $Y_{tt}=0.67(\xi_t\simeq 1)$ and $\D a_{\m}>50 \times 10^{-11}$ as an example, the mass of scalar
$m_{h}$ lies in a narrow window of $\sim 100 \leq m_{h}\leq 200$ GeV.

\section{Lepton flavor violation processes and the texture of Yukawa matrix}
The flavor changing Yukawa couplings will unavoidably lead to the 
enhancement of decay rates of  lepton flavor violating processes.
such as $\m\to e\g$, $\tau\to \m(e)\g$, $\m\to e^{-}e^{-}e^{+}$ and
$\t\to e^{-}e^{-}e^{+}(\m^{-}\m^{-}\m^{+}) $. The current experimental
data especially the data of $\m\to e \g$ will  impose strongest  constraints on
the related Yukawa couplings.  From the current data the upper bound of
the decay $\m\to e\g$ is
$\G(\m\to e\g) \leq 3.6 \times 10^{-30} $ GeV \cite{pdg:2000}.  It constrains the
coupling $Y_{e \t(\m)}$ to be extremely small.
In the models with flavor changing scalar interactions,
The leading contributions to $\m\to e\g$ come from the one loop
flavor changing diagram, the two loop double scalar exchanging
diagram and the two loop flavor changing Barr-Zee diagrams.

\begin{figure}[htb]\begin{center}
\includegraphics[width=0.6\textwidth]{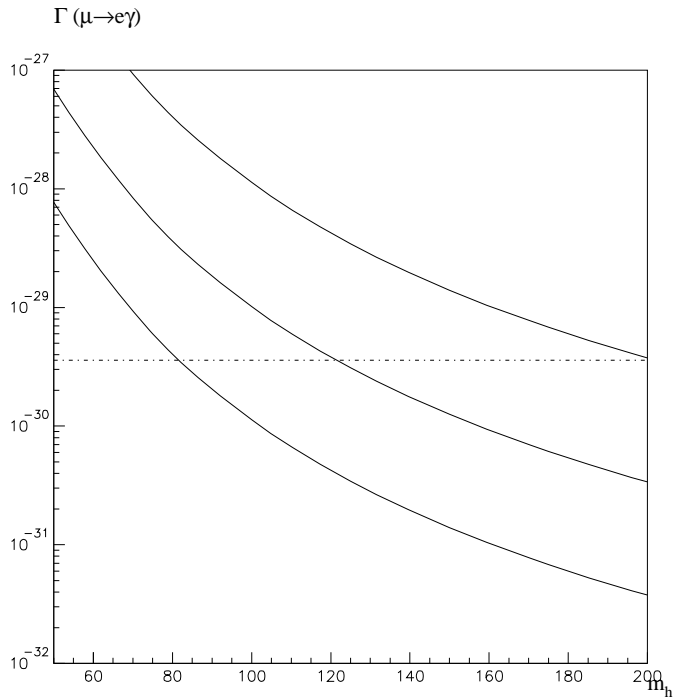} 
\caption{Contribution to decay $\m\to e\g$ from the sum of 
one loop and two loop double scalar diagrams. The three solid curves 
(from down to up) correspond to
 $Y_{\t e}=1\times 10^{-6}, 3\times 10^{-6}$ and $1\times 10^{-5}$ respectively. 
The coupling $Y_{\m\t}$ is taken to be 0.08.  The horizontal line indicates
the experimental upper bound of $\m\to e\g$ }.
\label{onetwo-meg.eps}
\end{center}\end{figure}

The effective vertex for one loop flavor changing scalar interaction reads
\cite{Chang:1993kw} 
\begin{align}\label{muegamma1}
\Gamma^{one}_{\m}=\frac{1}{2(4\pi)^2}\frac{m_{\tau}}{m_{\phi}^2}
\left(\ln\frac{m_{\phi}^2}{m_{\tau}^2}-\frac32 \right) Y_{\mu\tau}Y_{\tau e} 
\bar\ell i\sigma_{\mu\nu}\ell q^{\nu}
\end{align}
while the one for  two loop  loop double scalar exchanging is 
\begin{align}\label{muegamma2}
\Gamma^{two}_{\m}=\frac{N_{C} m_{\tau} m_{f}^2}{32\pi^4 m_{\phi}^4}
\left( \ln\frac{m_{\phi}^2}{m_{\tau}^2}-\frac52 \right)
Y_{ff}^2 Y_{\mu\tau}^2 \bar\ell i\sigma_{\mu\nu}\ell q^{\nu}
\end{align}
In Fig.\ref{onetwo-meg.eps} the decay rate from the  
sum of the first two diagrams are presented as function of the scalar  mass. In the calculation we take the value of coupling
$Y_{tt}=0.67$ (or  $\x_{t}\simeq 1$). The value of $Y_{\m\t}$ is
taken to be 0.08(or $\x_{\m\t}\simeq 50$) which is the typical
allowed value from the current data on \gmt .  It can be seen from the
figure that the decay rate of $\m\to e\g$ constrain the value of $Y_{\t e}$ to be no
more than $10^{-6}\sim 10^{-5}$ for  $100 \leq m_{h} \leq
200$GeV.

\begin{figure}[htb]\begin{center}
\includegraphics[width=0.6\textwidth]{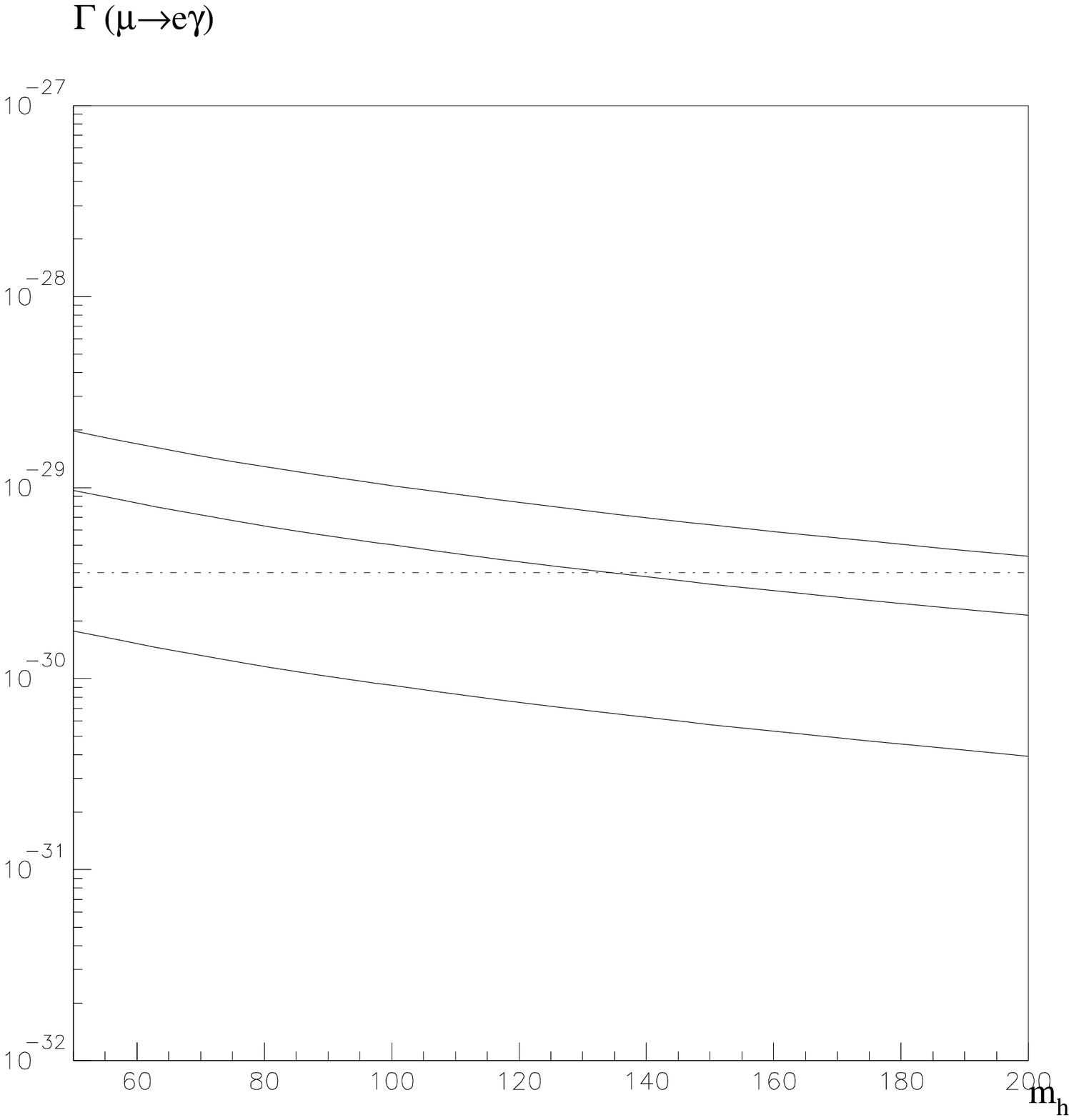} 
\caption{Contribution to decay $\m\to e\g$ from the two
loop flavor changing Barr-Zee diagrams. The three solid curves
(from down to up) correspond to
 $Y_{\m e}=3\times 10^{-6}, 7\times 10^{-6}$ and $1\times 10^{-5}$ respectively. 
  The horizontal line indicates the experimental upper bound of $\m\to e\g$}.
\label{barrzee-meg.eps}
\end{center}\end{figure}

Similarly, the value of coupling $Y_{\m e}$ is also constrained to be very
small by the decay rate $\m\to e\g$. The reason is that $Y_{\m e}$ is 
associated with the flavor changing Barr-Zee diagram in which muon goes
into tau in the lower loop. If there is no accidental cancellation
with other diagrams the upper bound of $Y_{\m e}$ can be obtained
by  assuming that the flavor changing Barr-Zee diagram is dominant.
The decay rate of $\mu\to e\gamma$ from this  diagram alone can be obtained
from Eq.(\ref{BarrZeeG2}) and is given by
\be\label{BarrZeemeg}
\Gamma^{BZ}(\mu\to e\gamma) &=& 8 \a m_{\mu}^5
\abs{{ N_c q^2_t  \over \pi^2 }
{m_\mu m_{t} \over m^2_{\phi}}  F\left({ m^2_{t}\over m^2_{\phi} }\right)
Y_{tt} Y_{\m\m}}^{2}
\label{BZ}
\ee
The numerical result is represented in Fig.\ref{barrzee-meg.eps} which shows that the 
 upper bound of $Y_{\m e}$ is also of the order $10^{-6}\sim 10^{-5}$ for $100 \leq m_{h} \leq 200$GeV.

With the above constraints on the values of  Yukawa couplings
in the lepton sector, let us discuss the possible texture of Yukawa   
coupling matrix. In the SM with one Higgs doublet, it is well known that
by assuming the Yukawa matrix to be of the Fritzsch form \cite{Fritzsch:1978vd,Fritzsch:1979zq}
in flavor basis, i.e.
\be
Y\simeq\left( 
\begin{array}{ccc}
      0   &  \sqrt{m_1 m_2}   &  0 \\
\sqrt{m_1 m_2}  &     0      & \sqrt{m_2 m_3} \\
     0     &     \sqrt{m_2 m_3}  &  m_3
\end{array}
\right)
\ee
one can reproduce not only correct quark masses in mass eigenstates
but also, in a good approximation, some of the mixing angles.  In the models with
multi-Higgs doublets, one can simply extend this Fritzsch
parameterization to all the other Yukawa matrices including the leptons\cite{cheng:1987rs}.
This results in the ansatz as in Eq.(\ref{Cheng-Sher_Ansartz}) with all $\xi_{ij}$
being of the same order of magnitude. 

It is not difficult to see that such an ansatz may be challenged by  current experiment data 
in the lepton sector. This is because in order to explain the possible large  muon \gmt, the off-diagonal
elements connecting  the second and third families   should be enhanced, while to meet the
constraints from $\m\to e \g$, the ones connecting the first and second or the first and third families
should be greatly suppressed.

Taking the value of $\Delta a_{\mu}> 50\times 10^{-11}$,  $m_h \sim
150 $GeV  and $m_A \gg m_h$ as an example , in the case of muon \gmt, if the
flavor-conserving Barr-Zee Diagram is playing the major role, the
rescaled coupling $\xi_{\m}$ should be as large as $50$ ( see
Fig.\ref{barrzee-g2.eps}).    If one assumes that
the flavor-changing coupling is responsible for the large muon \gmt,
$\xi_{\m\t}(Y_{\m\t})$ should be about 10(0.02).  On the other hand,
due to the strong constraint from $\m\to e\g$, for the flavor-changing
contribution dominated case, $\xi_{\t e}$ has to be less than 0.08
when $\xi_{\m\t}(Y_{\m\t})$ is taken a typical value of $17.6(0.03)$.  In the case of flavor
changing Barr-Zee diagram dominant, the Yukawa coupling $\xi_{\m e}$
has to be less than 0.24.  Thus one finds that 
\be
\xi_{\m} &\sim& \xi_{\m\t} \simeq {\mathcal O}(10) 
\nn
\xi_{\t e}&\sim& \xi_{\m e} \simeq {\mathcal O}(10^{-1})
\ee   
which clearly indicates that  the rescaled couplings $\xi_{ij}$are not in the same order of magnitude.
In the case of light pseudo-scalar mass $m_A \simeq 150$ GeV and $m_h \gg m_A$ the results
are similar.
%
%

%

From these  considerations, it is suggested that the Yukawa matrices associated with the physical
scalar bosons may take the following form in the mass eigenstate
\be
Y \simeq \l^2  \left(
\begin{array}{ccc}
     {\mathcal{O}}(1)    &     {\mathcal{O}}(\lambda^n)   &    {\mathcal{O}}(\lambda^n)  \\
      {\mathcal{O}}(\lambda^n)    &    {\mathcal{O}}(1)    &    {\mathcal{O}}(1) \\
       {\mathcal{O}}(\lambda^n)   &    {\mathcal{O}}(1)    &    {\mathcal{O}}(1)
\end{array}
\right)
\label{new-matrix}
\ee
where $\l\approx 0.22$ is roughly of the same order of  the Wolfenstein parameter $\l$, and $n\simeq 2\sim 3$.
With such a parameterization, one is able to understand all the current experimental data concerning
both muon \gmt and lepton flavor-changing processes.
   
\begin{figure}[htb]\begin{center}
\includegraphics[width=0.6\textwidth]{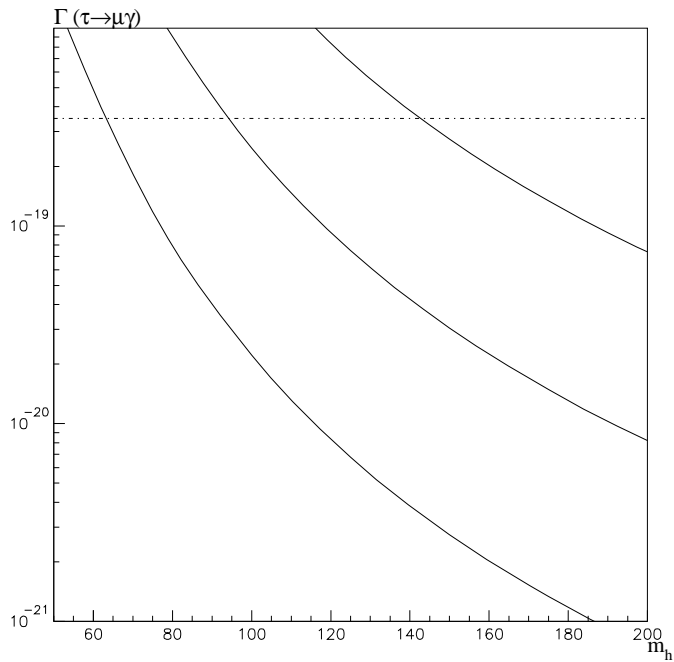} 
\caption{Prediction of  decay rate $\t\to \m\g$ from the sum of 
one loop and two loop double scalar diagrams. The three solid curves
(from down to up) correspond to
 $Y_{\t \t}$: 0.003, 0.01  and 0.03  respectively.
The coupling $Y_{\m\t}$ is taken to be 0.08.  The horizontal line indicates the experimental upper bound of $\t\to \m\g$}.
\label{onetwo-tmg.eps}
\end{center}\end{figure} 

If one takes the Yukawa matrix of the form in Eq.(\ref{new-matrix}), the decay rate
of $\t \to \m \g$ could be predicted.  In a good approximation, the decay rate can 
be obtained by replacing $Y_{\mu\tau}Y_{\tau e}$ into $Y_{\tau\tau} Y_{\tau \mu}$
in Eqs.(\ref{muegamma1}) and (\ref{muegamma2})  
Assuming $\t$ lepton dominance in the loop, the
 contributions to $\t \to \m\g$ 
and shown in Fig.\ref{onetwo-tmg.eps}.
The current upper bound on $\t\to\m \g$ is $3.5\times 10^{-19}$ GeV\cite{pdg:2000}.
It is found that,  the predicted decay rate could  reach the current experimental bound.
A modest improvement in the precision of the present experiment for $\t \to \m\g$ may
yield a first evidence of lepton family number non-conservation.


In summary, we have studied the muon \gmt and several lepton flavor
violation processes in the models with flavor-changing scalar
interactions.  The two loop diagrams with double scalar exchanges have
been investigated and their contribution to muon \gmt is found to be
compatible with the one from Barr-Zee diagram.  The constraints on
Yukawa coupling constants have been resulted from the current data of
muon \gmt and several lepton flavor violation processes.  The results
have shown a very strong constraints on the flavor-changing couplings
associated with the first generation lepton. The early ansatz that the
flavor changing couplings are proportional to the square root of the
products of related fermion masses may not be suitable for the lepton
sector if the $\Delta a_{\mu}$ is found to be $>50\times 10^{-11}$.
This indicates that both experimental and theoretical uncertainties need to be
further reduced in order to explore the existence of new physics from muon \gmt.  
It has been shown that an alternative simple parameterization
given in Eq.(\ref{new-matrix}) is more attractive to understand the current
experimental data.  With such a parameterization, the decay rate of
$\t\to \m\g$ is found to be close to the current experiment upper
bound.

\begin{acknowledgments}
This work is supported in part by the Chinese Academy of Sciences and NSFC under Grant
$\#$ 19625514. Y.F. Zhou acknowledges the  support by Alexander von Humboldt Foundation
\end{acknowledgments}

\begin{widetext}

\appendix

\section{Two loop double scalar  diagrams 
                 in muon \gmt and $\m\to e\g$} 
 From the  Yukawa interaction  shown in Eq.(\ref{Yukawa-interaction}).
 the  $\bar{q}q \phi$ vertex has the following form in $d$ dimension.
\be
  ig \m^{\e/2} ( Y_1+Y_2\g_{5} )
\ee
where $\m$ is renormalization scale and $\e/2=2-d/2$ . 
The total amplitude can be written as the product of lower
and upper parts as follows
\begin{align}
\Gamma_{\m}=M \cdot  I_{\m}
\end{align}
The amplitude $M$ for upper loop is given by
\begin{figure}[htb]\begin{center}
\includegraphics[width=0.6\textwidth]{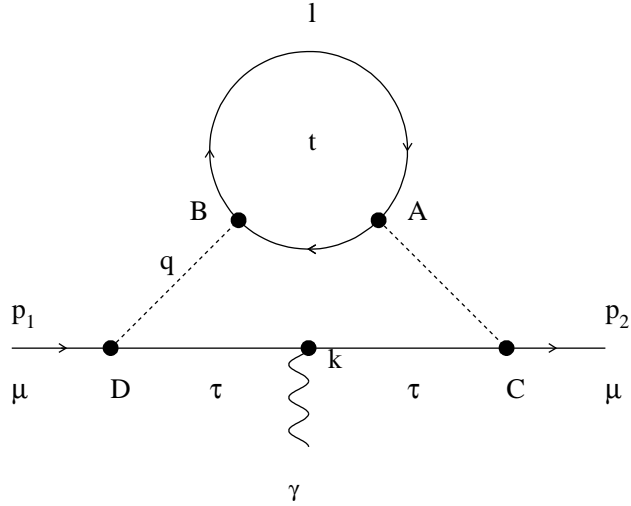}
\caption{ Two loop double scalar  exchanging digram }.
\label{2Ldh.eps}
\end{center}\end{figure}

\be
M&=&-g^{2}\m^{\e} \cdot 2N_{C}\mint{l}  \tr{{ ( A_{1}+A_{2}\gfive )
( \sh{l}+\not{q}+m_{f}) (B_{1}+B_{2}\gfive) {(\sh{l}+m_{f})}}}
{1\over (l+q)^2-m^2_\tau}{1\over l^2-m_{f}^2}
\nn
\ee
where $N_C$ and $m_f$ are the color number and mass of fermion $f$.
For $t$ quark $f=t$ and  $N_C=3$.  $A_{1,2}$ and $B_{1,2}$ are the couplings
of vertex $A$ and $B$.


The amplitude $I_{\mu}$ for the lower loop is given by
\be
I_{\m}
&=&-g^2 \m^{\e}  
    \bar{\ell} (p_2)(C_{1}+C_{2}\gfive)
     ( \not{p}_2-\not{q}+m_{\tau}) \gamma_{\mu} ( \not{p}_1-\not{q}+m_{\tau})
     (D_{1}+D_{2}\gfive)\ell(p_1) \nn
    &&\times{1\over (q-p_{2})^2- m_{\tau}^2} \cdot {1\over (q-p_{1})^2- m_{\tau}^2}\cdot {1\over (q^2-m_{\phi}^2)^2}
\ee
where $C_{1,2}$ and $D_{1,2}$ are the couplings for vertex $C$ and $D$.  


After integrating over the lower loop  and isolating the poles from Feynman integration, we obtain
\be
\G_{\m}
&=&{-8 \cdot 2N_{C} g^{4} (A_{1}B_{1}-A_{2}B_{2}) m_{\tau}m_{\mu} \over (4\p)^{4}}  \times \fint{x} 2 x(1-x) \fintb{y}{z} ( y+z )(1-y-z) 
\nn
&& \left(    
      \left( 
          ( { 2\over\e }-2\g_{E}+2\ln4\p - \ln x(1-x)+{ 1\over2 })\cdot f_{1,div}+2\cdot f_{1,con} \right)\right.
\nn
&&\left.  + { 1\over 2} 
C_{ab} R
   \cdot \left(
          ({ 2\over\e }-2\g_{E}+2\ln4\p - \ln x(1-x))\cdot f_{2,div}+2\cdot f_{2,con}
          \right)
\right)
\nn
&&    \bar{\ell} (C_{1}D_{1}+C_{2}D_{2}+(C_{1}D_{2}+C_{2}D_{1})\gfive ) 
        { i\sigma^{\mu\nu}k_{\nu}\over 2m_{\mu}}\ell
\label{total-AMP}
\ee
with $\D'=(y+z) m_{\t}^2+(1-y-z)) m_{\phi}^2$, $R={m_{f}^2 /[ x(1-x)]}$ 
and $C_{ab}={ 2A_{2}B_{2}/( A_{1}B_{1}-A_{2}B_{2}) }$.
In large $m_{f}$ limit, 
i.e. $m_{f}^2>> { 1\over 4}m_{\phi}^2>> m_{\t}^2$, 
the functions $f_{1,div(con)}$ and $f_{2,div(con)}$ have the following 
forms
\be
f_{1,div}&\to&{R\over \Delta'^2 },
\quad\quad
f_{2,div}\to-{ 1\over \D^{' 2} } 
\nn
f_{1,con} &\to& {R\over 2\Delta'^2 }
\left[
           1-  \ln({ \Delta' R\over \mu^4} ) 
\right],
\quad\quad
f_{2,con}\to { 1\over 2\D^{'2}} 
\left[
        1+ \ln{ \D' R\over \m^{4} }
\right]
\ee
After the renormalization in $\overline{\mbox{MS}}$ scheme for the upper loop, one finds
\be
\G_{\m}
&=&   { -8\cdot 2N_{C} g^{4}   m_{\t}m_{\m} (A_{1}B_{1}-A_{2}B_{2})\over (4\pi)^{4}} 
  \fint{x}  2x(1-x) \fintb{y}{z} (y+z) (1-y-z) 
\nn
&&
 \left[        +{3\D'+R\over \D'^2}
              ( -\ln x(1-x)+{ 1\over 2} +{\cal F}(x)+\ln{ \D'\over \m^{2} }) +2\cdot f_{1,con}
\right.
\nn
&&
\left.
        -{1\over \D'^2} { 1\over 2}C_{ab} R
        (  -\ln x(1-x)+{ 1\over 2}+{\cal F}(x)+\ln{ \D'\over \m^{2} })) 
         + { 1\over 2}C_{ab} R (2\cdot f_{2,con}-{ 1\over 2 })   
         +{3 \D'-2 R\over 2\D'^{2}}      
\right]
\nn
&&
        \bar{\ell}(p_{2})(C_{1}D_{1}+C_{2}D_{2}+(C_{1}D_{2}+C_{2}D_{1})\gfive))   
        { i\sigma_{\m\n}k^{\n} \over 2 m_{\m}  }\ell(p_{1})
\ee
with
\be
{\cal  F}(x)= \ln { m_{f}^2-x(1-x)m_{\t}^2\over \m^{2} }
\ee
In the limit of $m_{f}^2>>{ 1\over 4} m_{\phi}^2>>m_{\t}^2$, The 
above equation can be simplified as
\be
\G_{\m}&=&  - { N_{C} g^{4}   m_{\t}m_{\m} m_{f}^2(A_{1}B_{1}+A_{2}B_{2})\over 16\pi^{4}m_{\phi}^4} 
     \left( -{ 5\over 2}+\ln{ m_{\phi}^2\over m_{\t}^2 }\right)
\nn
&&
\left(
        (C_{1}D_{1}+C_{2}D_{2})  \bar{\ell}(p_{2}){ i\sigma_{\m\n}k^{\n} \over 2 m_{\m}  }\ell(p_{1}) 
       +(C_{1}D_{2}+C_{2}D_{1})  \bar{\ell}(p_{2}){ i\sigma_{\m\n}k^{\n}\gfive \over 2 m_{\m}  }\ell(p_{1}) 
\right) 
\ee
Therefore its contribution to muon \gmt is as follows
\be
\D a_{\m}= - { N_{C} g^{4}   m_{\t}m_{\m} m_{f}^2\over 16\pi^{4}m_{\phi}^4} 
     \left( -{ 5\over 2}+\ln{ m_{\phi}^2\over m_{\t}^2 }\right)
   (A_{1}B_{1}+A_{2}B_{2})         (C_{1}D_{1}+C_{2}D_{2})  
\ee
In the real coupling case, for scalar exchange, one has
\be
g A_{1}&=&g B_{1}=Y_{ff} ,\quad g C_{1}=g D_{1}=Y_{\m\t}
\ee
and others are zero. Similarly  for pseudoscalar exchange
the couplings are
\be
g A_{2}=g B_{2}= iY_{ff} , \quad g C_{2}=g D_{2}= iY_{\m\t}
 \ee
Therefore, the two loop double scalar (pseudo-scalar) diagram's contribution to $\D a_{\m}$ is
\be
\D a_{\m}
&=&
    \mp { N_{C}    m_{\t}m_{\m} m_{f}^2\over 16 \pi^{4}m_{\phi}^4} 
     \left( -{ 5\over 2}+\ln{ m_{\phi}^2\over m_{\t}^2 }\right)
     Y_{ff}^2 Y_{\m\t}^2
\ee
For  the decay $\m\to e \g$ ,  the effective vetex is
\be
\G^{(\mu\to e\gamma)}_{\m}&=&  - { N_{C} g^{4}   m_{\t}m_{\m} m_{f}^2(A_{1}B_{1}+A_{2}B_{2})\over 16\pi^{4}m_{\phi}^4} 
     \left( -{ 5\over 2}+\ln{ m_{\phi}^2\over m_{\t}^2 }\right)
\nn
&&
\left(
        (C'_{1}D_{1}+C'_{2}D_{2})  \bar{\ell}(p_{2}){ i\sigma_{\m\n}k^{\n} \over 2 m_{\m}  }\ell(p_{1}) 
       +(C'_{1}D_{2}+C'_{2}D_{1})  \bar{\ell}(p_{2}){ i\sigma_{\m\n}k^{\n}\gfive \over 2 m_{\m}  }\ell(p_{1}) 
\right) 
\nn
\ee
where $C'_{1}$ and $C'_{2}$ are the Yukawa couplings  for $\tau e \phi$ vetex.
The decay rate is then given by
\begin{align}
\Gamma(\mu\to e \gamma)=\frac{1}{16\pi m_{\mu}} \overline{\sum}
\abs{e \G^{(\mu\to e\gamma)}_{\mu} \e^{\mu}}^2
\end{align}
%

\end{widetext}

\bibliography{try,CoolRef,newRefs}   

\begin{thebibliography}{31}
\expandafter\ifx\csname natexlab\endcsname\relax\def\natexlab#1{#1}\fi
\expandafter\ifx\csname bibnamefont\endcsname\relax
  \def\bibnamefont#1{#1}\fi
\expandafter\ifx\csname bibfnamefont\endcsname\relax
  \def\bibfnamefont#1{#1}\fi
\expandafter\ifx\csname citenamefont\endcsname\relax
  \def\citenamefont#1{#1}\fi
\expandafter\ifx\csname url\endcsname\relax
  \def\url#1{\texttt{#1}}\fi
\expandafter\ifx\csname urlprefix\endcsname\relax\def\urlprefix{URL }\fi
\providecommand{\bibinfo}[2]{#2}
\providecommand{\eprint}[2][]{\url{#2}}

\bibitem[{\citenamefont{Bennett}(2002)}]{Bennett:2002jb}
\bibinfo{author}{\bibfnamefont{G.~W.} \bibnamefont{Bennett}}
  (\bibinfo{collaboration}{Muon g-2}) (\bibinfo{year}{2002}),
  \eprint[http://arXiv.org/abs]{hep-ex/0208001}.

\bibitem[{\citenamefont{Brown et~al.}(2001)}]{Brown:2001mg}
\bibinfo{author}{\bibfnamefont{H.~N.} \bibnamefont{Brown}} \bibnamefont{et~al.}
  (\bibinfo{collaboration}{Muon g-2}), \bibinfo{journal}{Phys. Rev. Lett.}
  \textbf{\bibinfo{volume}{86}}, \bibinfo{pages}{2227} (\bibinfo{year}{2001}),
  \eprint{hep-ex/0102017}.

\bibitem[{FJ0()}]{FJ02}
\bibinfo{note}{F. Jegerlehner, talk given at the Workshop Centre de Physique
  Theorique Marseille, France, 14-16 March 2002.}

\bibitem[{HMN()}]{HMNT}
\bibinfo{note}{K. Hagiwara, A. D. Martin, D. Nomura and T.Teubner, talk given
  by T.Teubner at ICHEP'02 , Amsterdam, Netherlands, 24-31 July 2002.}

\bibitem[{mod()}]{models-g2}
\bibinfo{note}{See for example, U.~Chattopadhyay, D.~K.~Ghosh and S.~Roy,Phys.\
  Rev.\ D {\bf 62}, 115001 (2000). L.~L.~Everett, G.~L.~Kane, S.~Rigolin and
  L.~T.~Wang, Phys.\ Rev.\ Lett.\ {\bf 86}, 3484 (2001). J.~L.~Feng and
  K.~T.~Matchev, Phys.\ Rev.\ Lett.\ {\bf 86}, 3480 (2001). U.~Chattopadhyay
  and P.~Nath, Phys.\ Rev.\ Lett.\ {\bf 86}, 5854 (2001). S.~Komine, T.~Moroi
  and M.~Yamaguchi, Phys.\ Lett.\ B {\bf 506}, 93 (2001). J.~Hisano and
  K.~Tobe, Phys.\ Lett.\ B {\bf 510}, 197 (2001). J.~E.~Kim, B.~Kyae and
  H.~M.~Lee, Phys.\ Lett.\ B {\bf 520}, 298 (2001). S.~P.~Martin and
  J.~D.~Wells, Phys.\ Rev.\ D {\bf 64}, 035003 (2001). H.~Baer, C.~Balazs,
  J.~Ferrandis and X.~Tata, Phys.\ Rev.\ D {\bf 64}, 035004 (2001).
  G.~C.~McLaughlin and J.~N.~Ng, Phys.\ Lett.\ B {\bf 493}, 88 (2000).
  S.~C.~Park and H.~S.~Song, Phys.\ Lett.\ B {\bf 506}, 99 (2001). C.~S.~Kim,
  J.~D.~Kim and J.~Song, Phys.\ Lett.\ B {\bf 511}, 251 (2001). K.~Agashe,
  N.~G.~Deshpande and G.~H.~Wu, Phys.\ Lett.\ B {\bf 511}, 85 (2001). K.~Choi,
  K.~Hwang, S.~K.~Kang, K.~Y.~Lee and W.~Y.~Song, Phys.\ Rev.\ D {\bf 64},
  055001 (2001). E.~Ma and M.~Raidal, Phys.\ Rev.\ Lett.\ {\bf 87}, 011802
  (2001). [Erratum-ibid.\ {\bf 87}, 159901 (2001)] R.~Arnowitt, B.~Dutta, B.~Hu
  and Y.~Santoso, Phys.\ Lett.\ B {\bf 505}, 177 (2001). V.~D.~Barger, T.~Falk,
  T.~Han, J.~Jiang, T.~Li and T.~Plehn, Phys.\ Rev.\ D {\bf 64}, 056007 (2001).
  R.~Casadio, A.~Gruppuso and G.~Venturi, Phys.\ Lett.\ B {\bf 495}, 378
  (2000). C.~H.~Chen and C.~Q.~Geng, Phys.\ Lett.\ B {\bf 511}, 77 (2001).
  Z.~z.~Xing, Phys.\ Rev.\ D {\bf 64}, 017304 (2001).}

\bibitem[{\citenamefont{Wu and Zhou}(2001)}]{wu:2001vq}
\bibinfo{author}{\bibfnamefont{Y.-L.} \bibnamefont{Wu}} \bibnamefont{and}
  \bibinfo{author}{\bibfnamefont{Y.-F.} \bibnamefont{Zhou}},
  \bibinfo{journal}{Phys. Rev} \textbf{\bibinfo{volume}{D64}},
  \bibinfo{pages}{115018} (\bibinfo{year}{2001}), \eprint{hep-ph/0104056}.

\bibitem[{\citenamefont{Raidal}(2001)}]{Raidal:2001pf}
\bibinfo{author}{\bibfnamefont{M.}~\bibnamefont{Raidal}},
  \bibinfo{journal}{Phys. Lett.} \textbf{\bibinfo{volume}{B508}},
  \bibinfo{pages}{51} (\bibinfo{year}{2001}), \eprint{arXiv:hep-ph/0103224}.

\bibitem[{wu:()}]{wu:S2HDM}
\bibinfo{note}{Y.L. Wu and L. Wolfenstein, Phys. Rev. Lett. {\bf 73}, 1762
  (1994). L. Wolfenstein and Y.L. Wu, Phys. Rev. Lett., {\bf 73}, 2809(1994).
  For more detailed analyses, see: Y.L. Wu, Carnegie-Mellon Report,
  hep-ph/9404241, 1994 (unpublished); {\it A Model for the Origin and
  Mechanisms of CP Violation}, in: Proceedings at 5th Conference on the
  Intersections of Particle and Nuclear Physics, St. Petersburg, FL, 31 May- 6
  Jun 1994, pp338, edited by S.J. Seestrom (AIP, New York, 1995).}

\bibitem[{\citenamefont{Lee}(1973)}]{Lee:1973iz}
\bibinfo{author}{\bibfnamefont{T.~D.} \bibnamefont{Lee}},
  \bibinfo{journal}{Phys. Rev.} \textbf{\bibinfo{volume}{D8}},
  \bibinfo{pages}{1226} (\bibinfo{year}{1973}).

\bibitem[{\citenamefont{Lee}(1974)}]{Lee:1974jb}
\bibinfo{author}{\bibfnamefont{T.~D.} \bibnamefont{Lee}},
  \bibinfo{journal}{Phys. Rept.} \textbf{\bibinfo{volume}{9}},
  \bibinfo{pages}{143} (\bibinfo{year}{1974}).

\bibitem[{2HD()}]{2HDM-type12}
\bibinfo{note}{P. Sikivie, Phys. Lett. {\bf B65}, 141 (1976); H.E. Haber, G.L.
  Kane and T. Sterling, Nucl. Phys. {\bf B161}, 493 (1979); N.G. Deshpande and
  E. Ma, Phys. rev. {\bf D18}, 2574 (1978); H. Georgi, Hadronic J. {\bf 1}, 155
  (1978); J.F. Donoghue and L.-F. Li, Phys. Rev. {\bf D19}, 945 (1979); A.B.
  Lahanas and C.E. Vayonakis, Phys. Rev. {\bf D19}, 2158 (1979); L.F. Abbott,
  P. Sikivie and M.B. Wise, Phys. Rev. {\bf D21}, 1393 (1980); G.C. Branco,
  A.J. Buras and J.M. Gerard, Nucl. Phys. {\bf B259}, 306 (1985); B. McWilliams
  and L.-F. Li, Nucl. Phys. {\bf B179}, 62 (1981); J.F. Gunion and H.E. Haber,
  Nucl. Phys. {\bf B272}, 1 (1986); J. Liu and L. Wolfenstein, Nucl. Phys. {\bf
  B289}, 1 (1987).}

\bibitem[{\citenamefont{Cheng and Sher}(1987)}]{cheng:1987rs}
\bibinfo{author}{\bibfnamefont{T.~P.} \bibnamefont{Cheng}} \bibnamefont{and}
  \bibinfo{author}{\bibfnamefont{M.}~\bibnamefont{Sher}},
  \bibinfo{journal}{Phys. Rev.} \textbf{\bibinfo{volume}{D35}},
  \bibinfo{pages}{3484} (\bibinfo{year}{1987}).

\bibitem[{\citenamefont{Sher and Yuan}(1991)}]{sher:1991km}
\bibinfo{author}{\bibfnamefont{M.}~\bibnamefont{Sher}} \bibnamefont{and}
  \bibinfo{author}{\bibfnamefont{Y.}~\bibnamefont{Yuan}},
  \bibinfo{journal}{Phys. Rev.} \textbf{\bibinfo{volume}{D44}},
  \bibinfo{pages}{1461} (\bibinfo{year}{1991}).

\bibitem[{\citenamefont{Dedes and Haber}(2001{\natexlab{a}})}]{Dedes:2001hh}
\bibinfo{author}{\bibfnamefont{A.}~\bibnamefont{Dedes}} \bibnamefont{and}
  \bibinfo{author}{\bibfnamefont{H.~E.} \bibnamefont{Haber}}
  (\bibinfo{year}{2001}{\natexlab{a}}), \eprint{hep-ph/0105014}.

\bibitem[{\citenamefont{Dedes and Haber}(2001{\natexlab{b}})}]{Dedes:2001nx}
\bibinfo{author}{\bibfnamefont{A.}~\bibnamefont{Dedes}} \bibnamefont{and}
  \bibinfo{author}{\bibfnamefont{H.~E.} \bibnamefont{Haber}},
  \bibinfo{journal}{JHEP} \textbf{\bibinfo{volume}{05}}, \bibinfo{pages}{006}
  (\bibinfo{year}{2001}{\natexlab{b}}), \eprint{hep-ph/0102297}.

\bibitem[{\citenamefont{Krawczyk}(2001)}]{Krawczyk:2001pe}
\bibinfo{author}{\bibfnamefont{M.}~\bibnamefont{Krawczyk}}
  (\bibinfo{year}{2001}), \eprint{hep-ph/0103223}.

\bibitem[{one()}]{oneloop-g2}
\bibinfo{note}{J. R. Primack and H.R. Quinn, Phys.~Rev. D{\bf 6}, 3171 (1972);
  W. A. Bardeen, R. Gastmans and B. Lautrup, Nucl.~Phys. B {\bf 46}, 319
  (1972); J. P. Leveille, Nucl.~Phys. B {\bf 137}, 63 (1978); H. E. Haber, G.
  L. Kane and T. Sterling, Nucl.~Phys. B {\bf 161}, 493 (1979); E. D. Carlson,
  S.L. Glashow and U. Sarid, Nucl.~Phys. B {\bf 309}, 597 (1988);}.

\bibitem[{Bar()}]{BarrZee}
\bibinfo{note}{J. D. Bjorken and S. Weinberg. \PRL{38}{622}{1977}, S. M. Barr
  and A. Zee, \PRL{65}{21}{1990}}.

\bibitem[{\citenamefont{Chang et~al.}(2001)\citenamefont{Chang, Chang, Chou,
  and Keung}}]{Chang:2000ii}
\bibinfo{author}{\bibfnamefont{D.}~\bibnamefont{Chang}},
  \bibinfo{author}{\bibfnamefont{W.-F.} \bibnamefont{Chang}},
  \bibinfo{author}{\bibfnamefont{C.-H.} \bibnamefont{Chou}}, \bibnamefont{and}
  \bibinfo{author}{\bibfnamefont{W.-Y.} \bibnamefont{Keung}},
  \bibinfo{journal}{Phys. Rev.} \textbf{\bibinfo{volume}{D63}},
  \bibinfo{pages}{091301} (\bibinfo{year}{2001}), \eprint{hep-ph/0009292}.

\bibitem[{\citenamefont{Cheung et~al.}(2001)\citenamefont{Cheung, Chou, and
  Kong}}]{Cheung:2001hz}
\bibinfo{author}{\bibfnamefont{K.}~\bibnamefont{Cheung}},
  \bibinfo{author}{\bibfnamefont{C.-H.} \bibnamefont{Chou}}, \bibnamefont{and}
  \bibinfo{author}{\bibfnamefont{O.~C.~W.} \bibnamefont{Kong}}
  (\bibinfo{year}{2001}), \eprint{hep-ph/0103183}.

\bibitem[{\citenamefont{Atwood et~al.}(1995)\citenamefont{Atwood, Reina, and
  Soni}}]{Atwood:1995ej}
\bibinfo{author}{\bibfnamefont{D.}~\bibnamefont{Atwood}},
  \bibinfo{author}{\bibfnamefont{L.}~\bibnamefont{Reina}}, \bibnamefont{and}
  \bibinfo{author}{\bibfnamefont{A.}~\bibnamefont{Soni}},
  \bibinfo{journal}{Phys. Rev. Lett.} \textbf{\bibinfo{volume}{75}},
  \bibinfo{pages}{3800} (\bibinfo{year}{1995}), \eprint{hep-ph/9507416}.

\bibitem[{\citenamefont{Atwood et~al.}(1997)\citenamefont{Atwood, Reina, and
  Soni}}]{Atwood:1997vj}
\bibinfo{author}{\bibfnamefont{D.}~\bibnamefont{Atwood}},
  \bibinfo{author}{\bibfnamefont{L.}~\bibnamefont{Reina}}, \bibnamefont{and}
  \bibinfo{author}{\bibfnamefont{A.}~\bibnamefont{Soni}},
  \bibinfo{journal}{Phys. Rev.} \textbf{\bibinfo{volume}{D55}},
  \bibinfo{pages}{3156} (\bibinfo{year}{1997}), \eprint{hep-ph/9609279}.

\bibitem[{\citenamefont{Wu and Zhou}(2000)}]{wu:1999fe}
\bibinfo{author}{\bibfnamefont{Y.~L.} \bibnamefont{Wu}} \bibnamefont{and}
  \bibinfo{author}{\bibfnamefont{Y.~F.} \bibnamefont{Zhou}},
  \bibinfo{journal}{Phys. Rev.} \textbf{\bibinfo{volume}{D61}},
  \bibinfo{pages}{096001} (\bibinfo{year}{2000}), \eprint{hep-ph/9906313}.

\bibitem[{\citenamefont{Sher}(2000)}]{sher:2000uq}
\bibinfo{author}{\bibfnamefont{M.}~\bibnamefont{Sher}}, \bibinfo{journal}{Phys.
  Lett.} \textbf{\bibinfo{volume}{B487}}, \bibinfo{pages}{151}
  (\bibinfo{year}{2000}), \eprint{hep-ph/0006159}.

\bibitem[{\citenamefont{Nie and Sher}(1998)}]{nie:1998dg}
\bibinfo{author}{\bibfnamefont{S.}~\bibnamefont{Nie}} \bibnamefont{and}
  \bibinfo{author}{\bibfnamefont{M.}~\bibnamefont{Sher}},
  \bibinfo{journal}{Phys. Rev.} \textbf{\bibinfo{volume}{D58}},
  \bibinfo{pages}{097701} (\bibinfo{year}{1998}), \eprint{hep-ph/9805376}.

\bibitem[{\citenamefont{Kang and Lee}(2001)}]{Kang:2001sq}
\bibinfo{author}{\bibfnamefont{S.~K.} \bibnamefont{Kang}} \bibnamefont{and}
  \bibinfo{author}{\bibfnamefont{K.~Y.} \bibnamefont{Lee}}
  (\bibinfo{year}{2001}), \eprint{hep-ph/0103064}.

\bibitem[{\citenamefont{Diaz et~al.}(2000)\citenamefont{Diaz, Martinez, and
  Rodriguez}}]{Diaz:2000cm}
\bibinfo{author}{\bibfnamefont{R.}~\bibnamefont{Diaz}},
  \bibinfo{author}{\bibfnamefont{R.}~\bibnamefont{Martinez}}, \bibnamefont{and}
  \bibinfo{author}{\bibfnamefont{J.~A.} \bibnamefont{Rodriguez}}
  (\bibinfo{year}{2000}), \eprint{hep-ph/0010149}.

\bibitem[{\citenamefont{Groom et~al.}(2000)}]{pdg:2000}
\bibinfo{author}{\bibfnamefont{D.~E.} \bibnamefont{Groom}} \bibnamefont{et~al.}
  (\bibinfo{collaboration}{Particle Data Group}), \bibinfo{journal}{EPJ.}
  \textbf{\bibinfo{volume}{C15}}, \bibinfo{pages}{1} (\bibinfo{year}{2000}).

\bibitem[{\citenamefont{Chang et~al.}(1993)\citenamefont{Chang, Hou, and
  Keung}}]{Chang:1993kw}
\bibinfo{author}{\bibfnamefont{D.}~\bibnamefont{Chang}},
  \bibinfo{author}{\bibfnamefont{W.~S.} \bibnamefont{Hou}}, \bibnamefont{and}
  \bibinfo{author}{\bibfnamefont{W.~Y.} \bibnamefont{Keung}},
  \bibinfo{journal}{Phys. Rev.} \textbf{\bibinfo{volume}{D48}},
  \bibinfo{pages}{217} (\bibinfo{year}{1993}), \eprint{hep-ph/9302267}.

\bibitem[{\citenamefont{Fritzsch}(1979)}]{Fritzsch:1979zq}
\bibinfo{author}{\bibfnamefont{H.}~\bibnamefont{Fritzsch}},
  \bibinfo{journal}{Nucl. Phys.} \textbf{\bibinfo{volume}{B155}},
  \bibinfo{pages}{189} (\bibinfo{year}{1979}).

\bibitem[{\citenamefont{Fritzsch}(1978)}]{Fritzsch:1978vd}
\bibinfo{author}{\bibfnamefont{H.}~\bibnamefont{Fritzsch}},
  \bibinfo{journal}{Phys. Lett.} \textbf{\bibinfo{volume}{B73}},
  \bibinfo{pages}{317} (\bibinfo{year}{1978}).

\end{thebibliography}
\bibliographystyle{apsrev}


\end{document}